\documentclass[12pt]{article}
\usepackage{latexsym}
\usepackage{graphicx}
\usepackage{amsmath}
\usepackage{amsfonts}
\usepackage{amssymb}
\usepackage{color}

\newcommand\efface[1]{}

\newtheorem{theorem}{Theorem}
\newtheorem{proposition}[theorem]{Proposition}
\newtheorem{lemma}[theorem]{Lemma}
\newtheorem{corollary}[theorem]{Corollary}

\newtheorem{example}[theorem]{Example}
\newtheorem{remark}[theorem]{Remark}

\newenvironment{proof}{
\par
\noindent {\bf Proof.}\rm}{\mbox{}\hfill\rule{0.5em}{0.809em}\par}

\newcommand{\hl}{\rule[3pt]{9pt}{0.5pt} \,\,} 
\newcommand{\vl}{\vrule height 6pt width 0.4pt} 

\begin{document}

\title{The Incidence Chromatic Number of Toroidal Grids}

\author{\'Eric Sopena\thanks{E-mail: sopena@labri.fr, corresponding author.}$\ $ and
        Jiaojiao Wu\thanks{E-mail: wujj0007@yahoo.com.tw.
        This work
        has been done while the author was visiting the LaBRI, supported by
        a postdoctoral fellowship from Bordeaux~1 University.} \\
        \\
  {\small Univ. Bordeaux, LaBRI, UMR5800, F-33400 Talence}\\
  {\small CNRS, LaBRI, UMR5800, F-33400 Talence}
}
\date{\today}

\maketitle

\begin{abstract}
An incidence in a graph $G$ is a pair
$(v,e)$ with $v \in V(G)$ and $e \in E(G)$, such that $v$ and $e$
are incident. Two incidences $(v,e)$ and $(w,f)$ are adjacent if $v=w$, 
or $e=f$, or the edge $vw$ equals $e$ or $f$.
The incidence chromatic number of $G$ is the smallest $k$ for which there
exists a mapping from the set of incidences of $G$ to a set of $k$ colors
that assigns distinct colors to adjacent incidences.

In this paper, we prove that the incidence chromatic number
 of the toroidal grid $T_{m,n}=C_m\Box C_n$
equals 5 when $m,n \equiv 0 \pmod 5$ and 6 otherwise.

\end{abstract}

\noindent{{\bf Key words: } Incidence coloring, Cartesian product of cycles, Toroidal grid.}

\noindent{{\bf 2000 Mathematics Subject Classification: } 05C15}

\section{Introduction}

Let $G$ be a graph with vertex set $V(G)$ and edge set $E(G)$.  
An {\em incidence} in $G$ is a pair
$(v,e)$ with $v \in V(G)$ and $e \in E(G)$, such that $v$ and $e$
are incident. We denote by $I(G)$ the set of all incidences in
$G$. Two incidences $(v,e)$ and $(w,f)$ are {\em adjacent} if one
of the following holds: (i) $v=w$, (ii) $e=f$, (iii) the edge
$vw$ equals $e$ or $f$.

An {\em incidence $k$-coloring} of $G$ is a mapping from $I(G)$
to a set of $k$ colors such that adjacent incidences are assigned
distinct colors. The {\em incidence chromatic number} $\chi_i(G)$ of $G$ is the
smallest $k$ such that $G$ admits an incidence $k$-coloring.

Incidence colorings were introduced by Brualdi and Massey in~\cite{BM}.
In that paper, the authors also conjectured that 
the relation $\chi_i(G)\le\Delta(G)+2$ holds for every graph $G$, where $\Delta(G)$ denotes
the maximum degree of $G$. In~\cite{G}, Guiduli disproved this
{\em Incidence Coloring Conjecture} (ICC for short).
Incidence coloring of various classes of graphs has been considered in the
litterature~\cite{G,HS,HSZ,HWC,LL,M,SLC,SS,WCP,W}
 and the ICC conjecture was proved to hold for several
classes such as trees, complete graphs and complete bipartite graphs~\cite{BM},
subcubic graphs~\cite{M}, 
$K_4$-minor free graphs~\cite{HSZ},
graphs with maximum average degree less than $\frac{22}{9}$~\cite{HS},
square, hexagonal and honeycomb meshes~\cite{HWC},
powers of paths~\cite{LL},
cubic Halin graphs~\cite{SS},
and Halin graphs with maximum degree at least 5~\cite{WCP}.
The problem of determining whether a given graph has incidence chromatic number at most $k$
or not was shown to be NP-complete by Li and Tu~\cite{LT}.

Incidence colorings are related to various types of vertex, edge or arc colorings.
For any graph $G$, let $H=H(G)$ be the bipartite graph given by 
$$V(H) = V(G) \cup E(G)$$
 and 
$$E(H) = \{(v,e):\ v \in V(G),\ e \in E(G),\ \mbox{$e$ and $v$ are incident in $G$}\}.$$
Each edge of $H$ corresponds to an incidence of $G$ and, therefore, any incidence coloring of $G$ 
corresponds to
a {\em strong edge coloring} (sometimes called a {\em distance-two edge-coloring})
of $H$, that is a proper coloring of the edges of $H$ such that each 
color class is an induced matching in $H$~\cite{EN}.

The {\em subdivision} $S(G)$ of $G$ is the graph obtained from $G$ by inserting a vertex 
of degree two on every edge of $G$. Any incidence coloring of $G$ then 
corresponds to a {\em distance-two vertex
coloring} of the line-graph $L(S(G))$ of $S(G)$, that is a vertex coloring such that any two vertices 
having the same color are at distance at least 3.

Let now $G^*$ be the digraph obtained from $G$ by replacing each edge of $G$ by two opposite arcs. 
Any incidence $(v,e)$ of $G$, with $e = vw$, can then be associated with the arc $vw$ in $G^*$. 
Therefore, any incidence coloring of $G$ corresponds to an arc-coloring of $G^*$ satisfying 
$(i)$ any two arcs having the same source vertex (of the form $uv$ and $uw$) are assigned 
distinct colors, $(ii)$ any two consecutive arcs (of the form $uv$ and $vw$) are assigned
 distinct colors. Hence, for every color $c$, the subgraph of $G^*$ induced by the $c$-colored arcs 
is a forest consisting of directed stars (whose arcs are directed towards the center). 
The incidence chromatic number of $G$ therefore equals the {\em directed star-arboricity}
 of $G^*$, as introduced by Algor and Alon in~\cite{AA}.

Let $G$ and $H$ be graphs. The {\em Cartesian product} $G \Box H$ of $G$ and $H$
is the graph with vertex set $V(G) \times V(H)$ where two vertices
$(u_1, v_1)$ and $(u_2,v_2)$ are adjacent if and only if either
$u_1=u_2$ and $v_1v_2 \in E(H)$, or $v_1=v_2$ and $u_1u_2 \in E(G)$.
Let $P_n$ and $C_n$ denote respectively the path and the cycle on $n$ vertices.
We will denote by $G_{m,n}=P_m\Box P_n$ the {\em grid} with $m$ rows and $n$ columns
and by $T_{m,n}=C_m\Box C_n$ the {\em toroidal grid} with $m$ rows and $n$ columns.

In this paper, we determine the incidence chromatic number of toroidal
grids and prove that this class of graphs satisfies the ICC:
\begin{theorem}
For every $m,n\geq 3$,
$\chi_i(T_{m,n})=5$ if $m,n \equiv 0 \pmod 5$ and
$\chi_i(T_{m,n})=6$ otherwise.
\label{SW-Theorem}
\end{theorem}

In~\cite{HWC}, Huang, Wang and Chung proved that
$\chi_i(G_{m,n})=5$ for every $m$, $n$. Since every toroidal grid $T_{m,n}$ contains 
the grid $G_{m,n}$ as a subgraph, we get that
$\chi_i(T_{m,n})\ge 5$ for every $m$, $n$.

The paper is organized as follows.
In Section~\ref{sec:2} we give basic properties and illustrate
the techniques we shall use in the proof of our main result, which
is given in Section~\ref{sec:3}.


\section{Preliminaries}
\label{sec:2}

Let $G$ be a graph, $u$ a vertex of $G$ with maximum
degree and $v$ a neighbour of $u$.
Since in any incidence coloring of $G$
all the incidences of the form $(u,e)$ have to get
distinct colors and all of them have to get
a color different from the color of $(v,vu)$, we
have:

\begin{proposition}
\label{prop-delta}
For every graph $G$, $\chi_i(G)\ge\Delta(G)+1$.
\end{proposition}

The {\em square} $G^2$ of a graph $G$ is given by
$V(G^2)=V(G)$ and $uv\in E(G^2)$ if and only if $uv\in E(G)$ or there exists $w\in V(G)$
such that $uw$, $vw\in E(G)$. In other words, any two vertices within distance at most
two in $G$ are linked by an edge in $G^2$.
Let now $c$ be a proper vertex coloring of $G^2$ and $\mu$
be the mapping defined by $\mu(u,uv)=c(v)$ for every incidence
$(u,uv)$ in $I(G)$. It is not difficult to check that $\mu$
is indeed an incidence coloring of $G$
(see Example~\ref{ex1} below). Therefore we have:

\begin{proposition}
\label{prop-square}
For every graph $G$, $\chi_i(G)\le\chi(G^2)$.
\end{proposition}

In~\cite{FGR}, Fertin, Goddard and Raspaud proved that the chromatic number of the square
of any $d$-dimensional grid $G_{n_1,\dots,n_d}$ is at most $2d+1$, which thus implies the 
above mentioned result concerning 2-dimensional grids~\cite{HWC}.

In~\cite{SW}, we studied the chromatic number of the squares
of toroidal grids and proved the following:

\begin{theorem}
Let $T_{m,n}=C_m \Box C_n$. Then  $\chi(T_{m,n}^2) \leq 7$ except that $\chi(T_{3,3}^2)=9$ and
$\chi(T_{3,5}^2)=\chi(T_{4,4}^2)=8$.
\label{SW-square-Theorem}
\end{theorem}

By Proposition~\ref{prop-square}, this result provides upper bounds on
the incidence chromatic number of toroidal grids.

In~\cite{SW}, we also proved the following:

\begin{theorem}
For every $m,n\ge 3$, $\chi(T_{m,n}^2) \ge 5$.
Moreover,  $\chi(T_{m,n}^2)=5$ if and only if 
$m,n\equiv 0 \pmod 5$.
\label{SW-square5-Theorem}
\end{theorem}

In~\cite{W}, the second author proved the following:

\begin{theorem}
For a regular graph $G$, $\chi_i(G)=\Delta(G)+1$
if and only if $\chi(G^2)=\Delta(G)+1$.
\label{W-theorem}
\end{theorem}

\begin{figure}
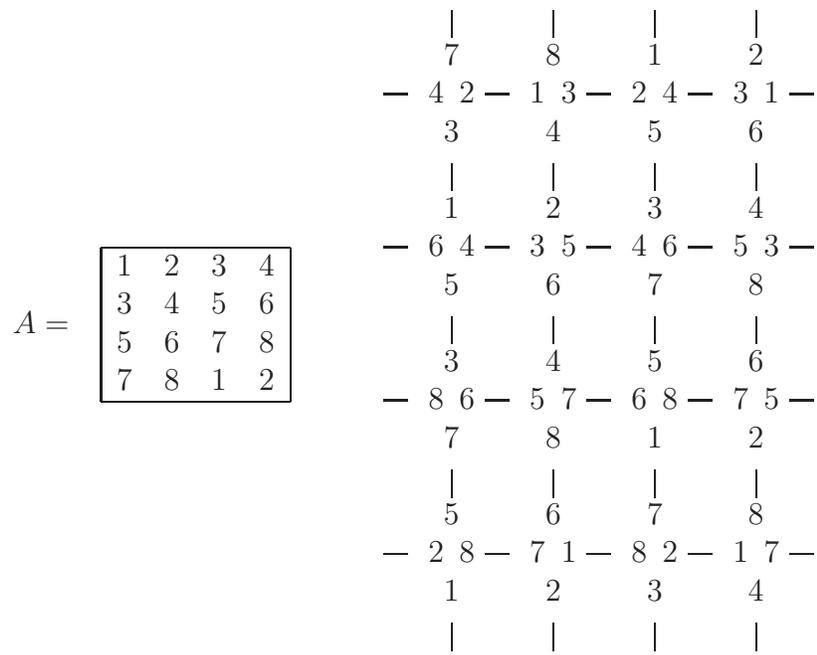

\begin{center}
\begin{tabular}{ccc}
$A=$
&
  \begin{tabular}{|cccc|}
     \hline
     1& 2& 3& 4 \\
     3& 4& 5& 6 \\
     5& 6& 7& 8 \\
     7& 8& 1& 2 \\
     \hline
  \end{tabular}
\ \ \ 
&
\ \ \ 
\begin{tabular}{@{}r@{}c@{}c@{}c@{}c@{}c@{}c@{}c@{}c@{}l}
 & \vl & & \vl & & \vl & & \vl & \\
 & 7 & & 8 & & 1 & & 2 \\
\hl 4 & & 2 \hl 1 & & 3 \hl 2 & & 4 \hl 3 & & 1 \hl \\
 & 3 & & 4 & & 5 & & 6  \\
& \vl & & \vl & & \vl & & \vl &  \\
 & 1 & & 2 & & 3 & & 4 \\
\hl 6 & & 4 \hl 3 & & 5 \hl 4 & & 6 \hl 5 & & 3 \hl \\
 & 5 & & 6 & & 7 & & 8 \\
  & \vl & & \vl & & \vl & & \vl & \\
 & 3 & & 4 & & 5 & & 6   \\
\hl 8 & & 6 \hl 5 & & 7 \hl 6 & & 8 \hl 7 & & 5 \hl   \\
 & 7 & & 8 & & 1 & & 2  \\
& \vl & & \vl & & \vl & & \vl &   \\
 & 5 & & 6 & & 7 & & 8   \\
\hl 2 & & 8 \hl 7 & & 1 \hl 8 & & 2 \hl 1 & & 7 \hl   \\
 & 1 & & 2 & & 3 & & 4  \\
& \vl & & \vl & & \vl & & \vl &   \\
 \end{tabular}
\end{tabular}
\caption{A pattern $A$ and the 
corresponding incidence coloring of $T_{4,4}$.}
\label{fig-pattern}
\end{center}
\end{figure}

Since toroidal grids are 4-regular, by 
combining Proposition~\ref{prop-delta},
Theorems~\ref{SW-square5-Theorem} and~\ref{W-theorem}
we get the following:

\begin{corollary}
For every $m,n\ge 3$, $\chi_i(T_{m,n})\ge 5$.
Moreover,  $\chi_i(T_{m,n})=5$ if and only if 
$m,n\equiv 0 \pmod 5$.
\label{square5-corollary}
\end{corollary}

Note here that this corollary is part of our main result.

Any vertex coloring of the square of a toroidal grid $T_{m,n}$ can be
given as an $m\times n$ matrix whose entries correspond in an obvious way
to the colors of the vertices.
Such a matrix will be called an $m\times n$ {\em pattern} in the following.

\begin{example}
\label{ex1}
{\rm Fig.~\ref{fig-pattern} shows a $4\times 4$ pattern $A$, which defines a vertex
coloring of $T_{4,4}^2$, and the incidence coloring of $T_{4,4}$
induced by this pattern, according to the discussion
before Proposition~\ref{prop-square}.
Note for instance that the four incidences of the form $(u,uv)$, for $u$
being the second vertex in the third row, have color 6, which corresponds
to the entry in row 3, column 2, of pattern $A$.
}
\end{example}

If $A$ and $B$ are patterns of size 
$m\times n$ and $m\times n'$ respectively, we shall denote by
$A+B$ the pattern of size $m\times (n+n')$ obtained by
``gluing'' together the patterns $A$ and $B$.
Moreover, we shall denote by $\ell A$, $\ell\ge 2$, the pattern
of size $m\times\ell n$ obtained by gluing
together $\ell$ copies of the pattern $A$.

We now shortly describe the technique we shall use
in the next section.
The main idea is to use a pattern for coloring the square
of a toroidal grid in order to get an incidence coloring
of this toroidal grid.
However, as shown in~\cite{SW}, the squares of  toroidal grids
are not all 6-colorable.
Therefore, we shall use the notion of a {\em quasi-pattern} which corresponds
to a vertex 6-coloring of the square of a {\em subgraph} of a toroidal grid
obtained by deleting some edges (namely those edges that cause a conflict when
transforming a vertex coloring to its corresponding incidence coloring).
We can then use such a quasi-pattern in the same way as before to obtain
a {\em partial} incidence coloring of the toroidal grid.
Finally, we shall prove that such a partial incidence coloring can
be extended to the whole toroidal grid without using any additional color
(most of the time, several distinct extensions are available and we shall
propose one of them).

We shall also use the following:
\begin{remark}
For every $m,n\ge 3$, $p,q\ge 1$, if $\chi_i(T_{m,n})\le k$
then $\chi_i(T_{pm,qn})\le k$.
\label{obs}
\end{remark}

To see that, it is enough to observe that every incidence $k$-coloring
$c$ of $T_{m,n}$ can be extended to an incidence $k$-coloring $c_{p,q}$ of
$T_{pm,qn}$ by ``repeating'' the pattern given by $c$, $p$ times ``vertically'' and $q$ times
``horizontally''.

\section{Proof of Theorem~1}
\label{sec:3}

According to Corollary~\ref{square5-corollary}, we only
need to prove that $\chi_i(T_{m,n})\le 6$ for every
$m,n\ge 3$. The proof is based on  a series of lemmas, according
to different values of $m$ and $n$.

We first consider the case when $m\equiv 0\pmod 3$.
We have proved in~\cite{SW} the following:

\begin{proposition}
If $k\ge 1$, $n\ge 3$ and $n$ even, then $\chi(T_{3k,n}^2)\le 6$.
\label{prop-3k}
\end{proposition}

Here we prove:

\begin{lemma}
If $k\ge 1$ and $n\ge 3$, then $\chi_i(T_{3k,n})\le 6$.
\label{lemma-3k}
\end{lemma}

\begin{figure}
$$\begin{array}{cccc}
  B=
  \begin{array}{|c|}
     \hline
     3\\
     1\\
     2\\
     \hline
  \end{array}

&
  C=
  \begin{array}{|cccc|}
     \hline
      1& 4 & 2 & 5\\
      2& 5 & 3 & 6\\
      3& 6 & 1 & 4\\
     \hline
  \end{array}
&
  D=
  \begin{array}{|cccc|}
     \hline
      1& 4 & 3 & 6\\
      2& 5 & 1 & 4\\
      3& 6 & 2 & 5\\
     \hline
  \end{array}
&
  E=
  \begin{array}{|cc|}
     \hline
      2 & 5 \\
      3 & 6 \\
      1 & 4 \\
     \hline
  \end{array}
\\
\end{array}$$
$$\begin{array}{cc}
  B+C=
  \begin{array}{|c|cccc|}
     \hline
     3& 1& 4 & 2 & 5\\
     1& 2& 5 & 3 & 6\\
     2& 3& 6 & 1 & 4\\
     \hline
  \end{array}

&
  B+D+E=
  \begin{array}{|c|cccc|cc|}
     \hline
     3& 1& 4 & 3 & 6& 2 & 5 \\
     1& 2& 5 & 1 & 4& 3 & 6 \\
     2& 3& 6 & 2 & 5& 1 & 4 \\
     \hline
  \end{array}
\\
\end{array}$$
\caption{Patterns and quasi-patterns for Lemma~\ref{lemma-3k}.}
\label{fig-pattern-1}
\end{figure}

\begin{figure}
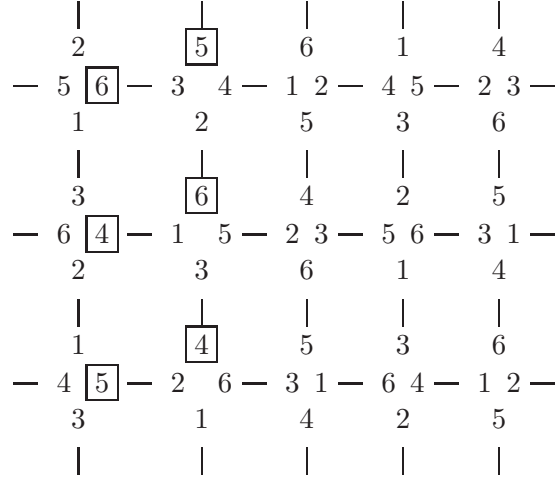

{\small
$$
\begin{tabular}{@{}r@{}c@{}c@{}c@{}c@{}c@{}c@{}c@{}c@{}c@{}c@{}c@{}l}
 & \vl & & \vl & & \vl & & \vl &  & \vl & \\
 & 2 & & \fbox{5} & & 6 & & 1 & & 4\\
\hl 5 & & \fbox{6} \hl 3 & & 4 \hl 1 & & 2 \hl 4 & & 5 \hl 2 & & 3 \hl \\
 & 1 & & 2 & & 5 & & 3 & & 6   \\
& \vl & & \vl & & \vl & & \vl &  & \vl & \\
 & 3 & & \fbox{6} & & 4 & & 2 & & 5 \\
\hl 6 & & \fbox{4} \hl 1 & & 5 \hl 2 & & 3 \hl 5 & & 6 \hl 3 & & 1 \hl  \\
 & 2 & & 3 & & 6 & & 1 & & 4\\
  & \vl & & \vl & & \vl & & \vl &  & \vl & \\
 & 1 & & \fbox{4} & & 5 & & 3 & & 6   \\
\hl 4 & & \fbox{5} \hl 2 & & 6 \hl 3 & & 1 \hl 6 & & 4 \hl 1 & & 2 \hl  \\
 & 3 & & 1 & & 4 & & 2 & & 5 \\
& \vl & & \vl & & \vl & & \vl &  & \vl & \\
 \end{tabular}$$
} 
\caption{Incidence coloring for Lemma~\ref{lemma-3k}\label{fig-lemma-1}.}
\end{figure}

\begin{proof}
If $n$ is even, the result follows from
Propositions~\ref{prop-square} and~\ref{prop-3k}.

We thus assume that $n$ is odd,
and we let first $k=1$. We consider three cases.
\begin{enumerate}

\item $n=3$.\\
We can easily get an incidence 6-coloring by coloring the incidences of one dimension with $\{1,2,3\}$ and the incidences of 
the other dimension with $\{4,5,6\}$.

\item $n=4\ell+1$.\\
Let $B$ and $C$ be the patterns depicted in Fig.~\ref{fig-pattern-1}
and consider the quasi-pattern $B+\ell C$ (the quasi-pattern $B+C$ 
is depicted in Fig.~\ref{fig-pattern-1}).
This quasi-pattern provides a $6$-coloring of $T_{m,n}^2$ if we delete all 
the edges linking vertices in the first column to
vertices in the second column.
We can use this quasi-pattern to obtain an incidence $6$-coloring of $T_{m,n}$
by modifying six incidence colors, as shown in Fig.~\ref{fig-lemma-1}
(modified colors are in boxes).

\item $n=4\ell+3$.\\
Let $B$, $D$ and $E$ be the patterns depicted in Fig.~\ref{fig-pattern-1}
and consider the quasi-pattern $B+\ell D+E$ (the quasi-pattern $B+D+E$ 
is depicted in Fig.~\ref{fig-pattern-1}).
As in the previous case,
we can use this quasi-pattern to obtain an incidence $6$-coloring of $T_{m,n}$
by modifying the same six incidence colors.
\end{enumerate}
For $k\ge 2$, the result now directly follows from Remark~\ref{obs}.
\end{proof}


We now consider the case when $m\equiv 0\pmod 4$.
For $m\equiv 0\pmod 5$, we have proved in~\cite{SW} the following:
\begin{proposition}
If $k\ge 1$, $n\ge 5$ and $n\neq 7$, then $\chi(T_{5k,n}^2)\le 6$.
\label{prop-5k}
\end{proposition}

Here we prove:

\begin{lemma}
If $k\ge 1$, $n\ge 3$ and $(k,n)\neq (1,5)$, then $\chi_i(T_{4k,n})\le 6$.
\label{lemma-4k}
\end{lemma}

\begin{figure}
$$ \begin{array}{cc}
F=\begin{array}{|ccc|}
    \hline
    1& 2& 4 \\
    1& 2& 4 \\
    3& 5& 6 \\
    3& 5& 6 \\
    \hline
     \end{array}
\
&
\

G=\begin{array}{|cccc|}
    \hline
    1& 2& 3& 4 \\
    1& 2& 3& 4 \\
    3& 4& 5& 6 \\
    3& 4& 5& 6 \\
    \hline
     \end{array}
\\
\end{array}
$$

$$H=2F+2G=\begin{array}{|ccc|ccc|cccc|cccc|}
    \hline
    1& 2& 4& 1& 2& 4& 1& 2& 3& 4& 1& 2& 3& 4 \\
    1& 2& 4& 1& 2& 4& 1& 2& 3& 4& 1& 2& 3& 4 \\
    3& 5& 6& 3& 5& 6& 3& 4& 5& 6& 3& 4& 5& 6 \\
    3& 5& 6& 3& 5& 6& 3& 4& 5& 6& 3& 4& 5& 6 \\
    \hline
     \end{array}$$\caption{Quasi-patterns for Lemma~\ref{lemma-4k}.}
\label{fig-pattern-2}
\end{figure}

\begin{figure}
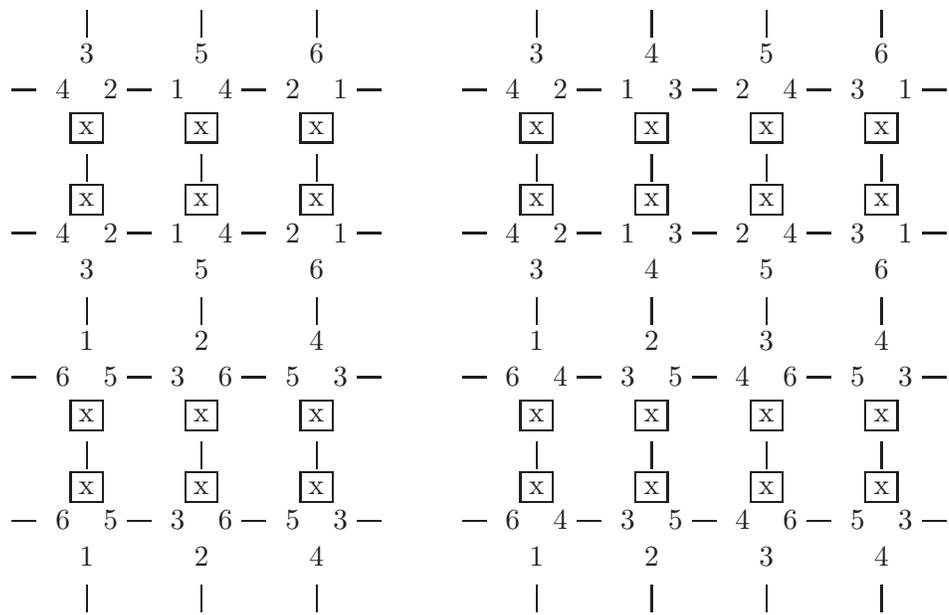

$$\begin{tabular}{cc}
{\small
\begin{tabular}{@{}r@{}c@{}c@{}c@{}c@{}c@{}c@{}c@{}c@{}c@{}l}
 & \vl & & \vl & & \vl &   \\
 & 3 & & 5 & & 6 &   \\
\hl 4 & & 2 \hl 1 & & 4 \hl 2 & & 1 \hl  \\
 & \fbox{x} & & \fbox{x} & & \fbox{x} &    \\
 & \vl & & \vl & & \vl &  \\
 & \fbox{x} & & \fbox{x} & & \fbox{x} &   \\
\hl 4 & & 2 \hl 1 & & 4 \hl 2 & & 1 \hl  \\
 & 3 & & 5 & & 6 &   \\
 & \vl & & \vl & & \vl &   \\
 & 1 & & 2 & & 4 &   \\
\hl 6 & & 5 \hl 3 & & 6 \hl 5 & & 3 \hl  \\
 & \fbox{x} & & \fbox{x} & & \fbox{x} &    \\
 & \vl & & \vl & & \vl &  \\
 & \fbox{x} & & \fbox{x} & & \fbox{x} & \\
\hl 6 & & 5 \hl 3 & & 6 \hl 5 & & 3 \hl  \\
 & 1 & & 2 & & 4 &  \\
 & \vl & & \vl & & \vl &  \\
\end{tabular}

} 
\
&
\
{\small

\begin{tabular}{@{}r@{}c@{}c@{}c@{}c@{}c@{}c@{}c@{}c@{}c@{}l}
 & \vl & & \vl & & \vl & & \vl &  \\
 & 3 & & 4 & & 5 & & 6 &  \\
\hl 4 & & 2 \hl 1 & & 3 \hl 2 & & 4 \hl 3 & & 1 \hl \\
 & \fbox{x} & & \fbox{x} & & \fbox{x} & & \fbox{x} &   \\
 & \vl & & \vl & & \vl & & \vl &  \\
 & \fbox{x} & & \fbox{x} & & \fbox{x} & & \fbox{x} &  \\
\hl 4 & & 2 \hl 1 & & 3 \hl 2 & & 4 \hl 3 & & 1 \hl \\
 & 3 & & 4 & & 5 & & 6 &   \\
 & \vl & & \vl & & \vl & & \vl &  \\
 & 1 & & 2 & & 3 & & 4 &  \\
\hl 6 & & 4 \hl 3 & & 5 \hl 4 & & 6 \hl 5 & & 3 \hl \\
 & \fbox{x} & & \fbox{x} & & \fbox{x} & & \fbox{x} &   \\
 & \vl & & \vl & & \vl & & \vl &  \\
 & \fbox{x} & & \fbox{x} & & \fbox{x} & & \fbox{x} & \\
\hl 6 & & 4 \hl 3 & & 5 \hl 4 & & 6 \hl 5 & & 3 \hl \\
 & 1 & & 2 & & 3 & & 4 &  \\
 & \vl & & \vl & & \vl & & \vl &  \\
\end{tabular}

} 

\\
\end{tabular}$$

\caption{Partial incidence colorings for Lemma~\ref{lemma-4k}\label{fig-lemma-2}.}
\end{figure}

\begin{proof}
For $n=5$, the result holds by Proposition~\ref{prop-5k}, except for $k=1$. 

Assume now $k=1$ and $n \neq 5$ and consider
the quasi-patterns $F$ and $G$ depicted
in Fig.~\ref{fig-pattern-2}. 
From these patterns, we can derive a partial incidence
6-coloring of $T_{4,3}$ and $T_{4,4}$, respectively,
as shown in Fig.~\ref{fig-lemma-2}, where the
uncolored incidences are denoted by $x$.
It is easy to check that every such incidence has
only four forbidden colors and that only incidences
belonging to a same edge have to be distinct.
Therefore, these partial incidence colorings can be
extended to incidence $6$-colorings of $T_{4,3}$ and $T_{4,4}$.

For $n\ge 6$, we shall use the quasi-pattern $H=pF+qG$
where $p$ and $q$ satisfy $n=3p+4q$ (recall that
every integer except 1,2 and 5 can be written in this form).
The quasi-pattern $H=2F+2G$ is depicted in Fig.~\ref{fig-pattern-2}.
As in the previous case, this quasi-pattern provides a partial
incidence $6$-coloring of $T_{4,n}$ that can be extended
to an incidence $6$-coloring of $T_{4,n}$.

For $k\ge 2$, the result now directly follows from Remark~\ref{obs}.
\end{proof}

\begin{figure}
$$ \begin{array}{cc}
I=\begin{array}{|cccccc|}
    \hline
    6& 1& 2& 3& 4& 5 \\
    3& 4& 5& 6& 1& 2 \\
    5& 6& 1& 2& 3& 4 \\
    2& 3& 4& 5& 6& 1 \\
    4& 5& 6& 1& 2& 3 \\
    \hline
     \end{array}
\
&
\
I'=\begin{array}{|cccccc|}
    \hline
    6& 1& 2& 3& 4& 5 \\
    6& 1& 2& 3& 4& 5 \\
    6& 1& 2& 3& 4& 5 \\
    3& 4& 5& 6& 1& 2 \\
    5& 6& 1& 2& 3& 4 \\
    2& 3& 4& 5& 6& 1 \\
    4& 5& 6& 1& 2& 3 \\
    \hline
     \end{array}
\end{array}$$
\caption{Patterns for Lemma~\ref{lemma-mn}.}
\label{fig-pattern-3}
\end{figure}

\begin{figure}
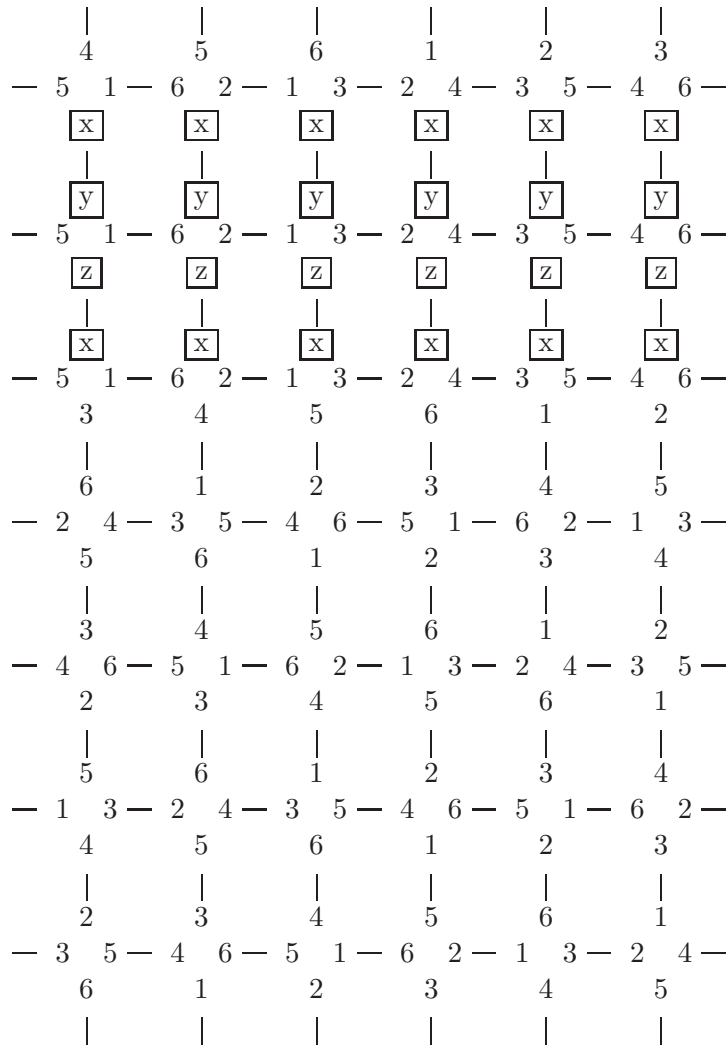

{\small
$$
\begin{tabular}{@{}r@{}c@{}c@{}c@{}c@{}c@{}c@{}c@{}c@{}c@{}c@{}c@{}l}
 & \vl & & \vl & & \vl & & \vl &  & \vl & & \vl &\\
 & 4 & & 5 & & 6 & & 1 & & 2 & & 3\\
 \hl 5 & & 1 \hl 6 & & 2 \hl 1 & & 3 \hl 2 & & 4 \hl 3 & & 5 \hl  4 & & 6 \hl\\
 & \fbox{x} & & \fbox{x} & & \fbox{x} & & \fbox{x} & & \fbox{x} & & \fbox{x}\\
 & \vl & & \vl & & \vl & & \vl &  & \vl & & \vl &\\
  & \fbox{y} & & \fbox{y} & & \fbox{y} & & \fbox{y} & & \fbox{y} & & \fbox{y}\\
 \hl 5 & & 1 \hl 6 & & 2 \hl 1 & & 3 \hl 2 & & 4 \hl 3 & & 5 \hl  4 & & 6 \hl\\
 & \fbox{z} & & \fbox{z} & & \fbox{z} & & \fbox{z} & & \fbox{z} & & \fbox{z}\\
  & \vl & & \vl & & \vl & & \vl &  & \vl & & \vl &\\
 & \fbox{x} & & \fbox{x} & & \fbox{x} & & \fbox{x} & & \fbox{x} & & \fbox{x}\\
 \hl 5 & & 1 \hl 6 & & 2 \hl 1 & & 3 \hl 2 & & 4 \hl 3 & & 5 \hl  4 & & 6 \hl\\
 & 3 & & 4 & & 5 & & 6 & & 1 & & 2\\
 & \vl & & \vl & & \vl & & \vl &  & \vl & & \vl &\\
  & 6 & & 1 & & 2 & & 3 & & 4 & & 5\\
\hl 2 & & 4 \hl 3 & & 5 \hl 4 & & 6 \hl 5 & & 1 \hl 6 & & 2 \hl  1 & & 3 \hl  \\
 & 5 & & 6 & & 1 & & 2 & & 3 & & 4\\
 & \vl & & \vl & & \vl & & \vl &  & \vl & & \vl &\\
 & 3 & & 4 & & 5 & & 6 & & 1 & & 2\\
 \hl 4 & & 6 \hl 5 & & 1 \hl 6 & & 2 \hl 1 & & 3 \hl 2 & & 4 \hl 3 & & 5 \hl  \\
   & 2 & & 3 & & 4 & & 5 & & 6 & & 1\\
  & \vl & & \vl & & \vl & & \vl &  & \vl & & \vl &\\
 & 5 & & 6 & & 1 & & 2 & & 3 & & 4\\
\hl 1 & & 3 \hl 2 & & 4 \hl 3 & & 5 \hl 4 & & 6 \hl 5 & & 1 \hl  6 & & 2 \hl  \\
 & 4 & & 5 & & 6 & & 1 & & 2 & & 3\\
 & \vl & & \vl & & \vl & & \vl &  & \vl & & \vl &\\
  & 2 & & 3 & & 4 & & 5 & & 6 & & 1\\
\hl 3 & & 5 \hl 4 & & 6 \hl 5 & & 1 \hl 6 & & 2 \hl 1 & & 3 \hl  2 & & 4 \hl  \\
 & 6 & & 1 & & 2 & & 3 & & 4 & & 5\\
 & \vl & & \vl & & \vl & & \vl &  & \vl & & \vl &\\
 \end{tabular}$$
} 
\caption{A partial incidence coloring of $T_{7,6}$.}
\label{fig-lemma-3}
\end{figure}

We now consider the remaining cases.

\begin{lemma}
If $m, n \geq 5$, $m \neq 6, 8$ and $n \neq 7$,  then $\chi_i(T_{m,n}) \leq 6$.
\label{lemma-mn}
\end{lemma}

\begin{proof}
Assume $m, n \geq 5$, $m \neq 6, 8$ and $n \neq 7$.
By Proposition~\ref{prop-5k}, we have
$\chi(T^2_{5k,n})\le 6$ for $n \neq 7$. Hence, there exists a vertex $6$-coloring
of $T_{5k,n}^2$ which corresponds to some pattern $M$ of size $5k\times n$.
We claim that each row of pattern $M$ can be repeated one or three times
to get quasi-patterns that can be extended to incidence 6-colorings of 
the corresponding toroidal grids.

Let for instance $M'$ be the quasi-pattern obtained from $M$ by repeating the first row
of $M$ three times. The quasi-pattern $M'$ has thus size $(5k+2)\times n$.
The quasi-pattern $M'$ induces a partial incidence coloring of $T_{5k+2,n}$
in which the only uncolored incidences are those lying on the edges
linking vertices in the first row to vertices in the second row and on the edges
linking vertices in the second row to vertices in the third row.

We illustrate this in Fig.~\ref{fig-pattern-3} with a pattern $I$
of size $5\times 6$ (this pattern induces a vertex 6-coloring
of $T_{5,6}^2$) and its associated pattern $I'$ of size $7\times 6$.
The partial incidence coloring of $T_{7,6}$ obtained from $I'$ is then
given in Fig.~\ref{fig-lemma-3}, where uncolored incidences are
denoted by $x$, $y$ and $z$.

Observe now that in each column, the two incidences denoted by $x$
have three forbidden colors in common and each of them has four forbidden colors in total.
Therefore, we can assign them the same color. Now, in each column,
the incidences denoted by $y$ and $z$ have four forbidden colors in common (the color
assigned to $x$ is one of them) and each of them has five forbidden colors
in total. They can be thus colored with distinct colors.
Doing that, we extend the partial incidence
coloring of $T_{7,6}$ to an incidence $6$-coloring of $T_{7,6}$.

The same technique can be used for obtaining an incidence 6-coloring
of $T_{5k+2,n}$ since all the columns are ``independent'' in the
quasi-pattern $M'$, with respect to uncolored incidences.

If we repeat three times several distinct rows of pattern $M$, each
repeated row will produce a chain of four uncolored incidences, as before,
and any two such chains in the same column will be ``independent'', since they will be
separated by an edge whose incidences
are both colored.
Hence, we will be able to extend the corresponding quasi-pattern to an
incidence 6-coloring of the toroidal grid, by assigning available colors
to each chain as we did above.

Starting from a pattern $M$ of size $5k\times n$, we can thus obtain
quasi-patterns of size $(5k+2)\times n$,
$(5k+4)\times n$,
$(5k+6)\times n$ and
$(5k+8)\times n$, by repeating respectively one, two, three or four lines from $M$.
Using these quasi-patterns, we can produce incidence 6-colorings of the
toroidal grid $T_{m,n}$, $m,n\ge 5$, $n\neq 7$, for every $m$
except $m=6$, $8$.
\end{proof}

\begin{figure}
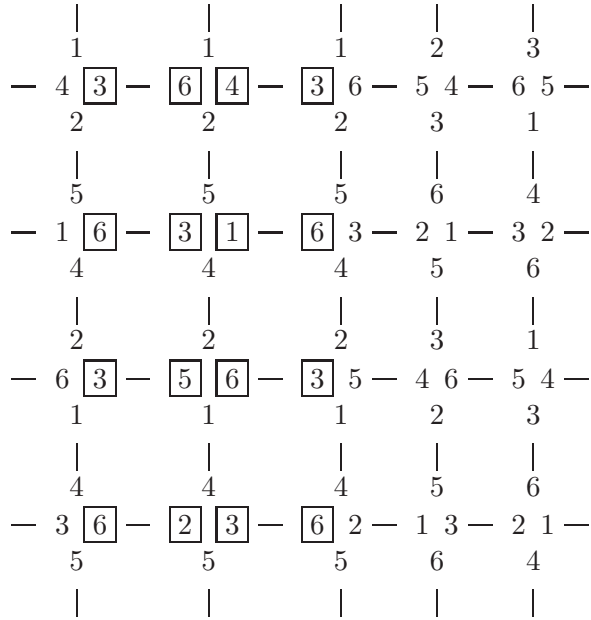

{\small
$$
\begin{tabular}{@{}r@{}c@{}c@{}c@{}c@{}c@{}c@{}c@{}c@{}c@{}l}
 & \vl & & \vl & & \vl & & \vl &  & \vl & \\
 & 1 & & 1 & & 1 & & 2 & & 3 &  \\
\hl 4 & & \fbox{3} \hl \fbox{6} & & \fbox{4} \hl \fbox{3} & & 6 \hl 5 & & 4 \hl 6 & & 5 \hl\\
 & 2 & & 2 & & 2 & & 3 & & 1 &  \\
 & \vl & & \vl & & \vl & & \vl &  & \vl & \\
 & 5 & & 5 & & 5 & & 6 & & 4 &  \\
\hl 1 & & \fbox{6} \hl \fbox{3} & & \fbox{1} \hl \fbox{6} & & 3 \hl 2 & & 1 \hl 3 & & 2 \hl\\
 & 4 & & 4 & & 4 & & 5 & & 6 &  \\
 & \vl & & \vl & & \vl & & \vl &  & \vl & \\
 & 2 & & 2 & & 2 & & 3 & & 1 &  \\
\hl 6 & & \fbox{3} \hl \fbox{5} & & \fbox{6} \hl \fbox{3} & & 5 \hl 4 & & 6 \hl 5 & & 4 \hl\\
 & 1 & & 1 & & 1 & & 2 & & 3 &  \\
 & \vl & & \vl & & \vl & & \vl &  & \vl & \\
 & 4 & & 4 & & 4 & & 5 & & 6 &  \\
\hl 3 & & \fbox{6} \hl \fbox{2} & & \fbox{3} \hl \fbox{6} & & 2 \hl 1 & & 3 \hl 2 & & 1 \hl\\
 & 5 & & 5 & & 5 & & 6 & & 4 &  \\
 & \vl & & \vl & & \vl & & \vl &  & \vl & \\
\end{tabular}$$
} 
\caption{An incidence 6-coloring of $T_{4,5}$.}
\label{fig-45}
\end{figure}

\begin{figure}
$$\begin{array}{cc}
J=
&
   \begin{array}{|ccccccc|}
     \hline
     3& 5& 6 & 3 & 4& 5 & 6 \\
     1& 2& 4 & 1 & 2& 3 & 4 \\
     1& 2& 4 & 1 & 2& 3 & 4 \\
     3& 5& 6 & 3 & 4& 5 & 6 \\
     3& 5& 6 & 3 & 4& 5 & 6 \\
     1& 2& 4 & 1 & 2& 3 & 4 \\
     1& 2& 4 & 1 & 2& 3 & 4 \\
     \hline
  \end{array}
\end{array}$$
\caption{A quasi-pattern for Lemma~\ref{lemma-77}.}
\label{fig-pattern-77}
\end{figure}

The only remaining cases are $m=4$, $n=5$ and $m=n=7$.
Then we have:

\begin{lemma}\label{lemma-77}
$\chi_i(T_{4,5}) \leq 6$ and $\chi_i(T_{7,7}) \leq 6$.
\end{lemma}

\begin{proof}
Let $m=4$ and $n=5$. Consider the pattern $C$ of size $3\times 4$ depicted
in Fig.~\ref{fig-pattern-1}.
As in the proof of Lemma~\ref{lemma-mn}, we can repeat the first row of $C$
three times to get a quasi-pattern $C'$ that can be extended to an incidence
6-coloring of $T_{5,4}$. We then exchange $m$ and $n$ to get an incidence
6-coloring of $T_{4,5}$, depicted in Fig.~\ref{fig-45} (the colors
assigned to uncolored incidences are drawn in boxes).

\begin{figure}
{\small
$$
\begin{tabular}{@{}r@{}c@{}c@{}c@{}c@{}c@{}c@{}c@{}c@{}c@{}c@{}c@{}c@{}c@{}l}
 & \vl & & \vl & & \vl & & \vl &  & \vl & & \vl & & \vl &\\
 & 1 & & 2 & & 4 & & 1 & & 2 & & 3 & & 4 & \\
\hl 6 & & \fbox{4} \hl 3 & & \fbox{1} \hl 5 & & \fbox{2} \hl 6 & & 4 \hl 3 & & 5 \hl 4 & & 6 \hl 5 & & 3 \hl\\
 & \fbox{5} & & \fbox{6} & & \fbox{3} & & \fbox{5} & & \fbox{6} & & \fbox{1} & & \fbox{2} &  \\
& \vl & & \vl & & \vl & & \vl &  & \vl & & \vl & & \vl &\\
 & 3 & & 5 & & 6 & & 3 & & 4 & & 5 & & 6 & \\
\hl 4 & & 2 \hl 1 & & 4 \hl 2 & & 1 \hl 4 & & 2 \hl 1 & & 3 \hl 2 & & 4 \hl 3 & & 1 \hl \\
 & \fbox{y} & & \fbox{y} & & \fbox{y} & & \fbox{y} & & \fbox{y} & & \fbox{y} & & \fbox{y} &  \\
  & \vl & & \vl & & \vl & & \vl &  & \vl & & \vl & & \vl &\\
 & \fbox{5} & & \fbox{6} & & \fbox{3} & & \fbox{5} & & \fbox{6} & & \fbox{1} & & \fbox{2} &  \\
\hl 4 & & 2 \hl 1 & & 4 \hl 2 & & 1 \hl 4 & & 2 \hl 1 & & 3 \hl 2 & & 4 \hl 3 & & 1 \hl \\
 & 3 & & 5 & & 6 & & 3 & & 4 & & 5 & & 6 & \\
& \vl & & \vl & & \vl & & \vl &  & \vl & & \vl & & \vl &\\
  & 1 & & 2 & & 4 & & 1 & & 2 & & 3 & & 4 & \\
\hl 6 & & 5 \hl 3 & & 6 \hl 5 & & 3 \hl 6 & & 4 \hl 3 & & 5 \hl 4 & & 6 \hl 5 & & 3 \hl\\
 & \fbox{x} & & \fbox{x} & & \fbox{x} & & \fbox{x} & & \fbox{x} & & \fbox{x} & & \fbox{x} &  \\
 & \vl & & \vl & & \vl & & \vl &  & \vl & & \vl & & \vl &\\
 & \fbox{x} & & \fbox{x} & & \fbox{x} & & \fbox{x} & & \fbox{x} & & \fbox{x} & & \fbox{x} &  \\
\hl 6 & & 5 \hl 3 & & 6 \hl 5 & & 3 \hl 6 & & 4 \hl 3 & & 5 \hl 4 & & 6 \hl 5 & & 3 \hl\\
 & 1 & & 2 & & 4 & & 1 & & 2 & & 3 & & 4 &  \\
& \vl & & \vl & & \vl & & \vl &  & \vl & & \vl & & \vl &\\
 & 3 & & 5 & & 6 & & 3 & & 4 & & 5 & & 6 & \\
\hl 4 & & 2 \hl 1 & & 4 \hl 2 & & 1 \hl 4 & & 2 \hl 1 & & 3 \hl 2 & & 4 \hl 3 & & 1 \hl \\
 & \fbox{x} & & \fbox{x} & & \fbox{x} & & \fbox{x} & & \fbox{x} & & \fbox{x} & & \fbox{x} &  \\
 & \vl & & \vl & & \vl & & \vl &  & \vl & & \vl & & \vl &\\
 & \fbox{x} & & \fbox{x} & & \fbox{x} & & \fbox{x} & & \fbox{x} & & \fbox{x} & & \fbox{x} &  \\
\hl 4 & & 2 \hl 1 & & 4 \hl 2 & & 1 \hl 4 & & 2 \hl 1 & & 3 \hl 2 & & 4 \hl 3 & & 1 \hl \\
 & 3 & & 5 & & 6 & & 3 & & 4 & & 5 & & 6 & \\
 & \vl & & \vl & & \vl & & \vl &  & \vl & & \vl & & \vl &\\
\end{tabular}$$
} 
\caption{A partial incidence 6-coloring of $T_{7,7}$.}
\label{fig-77}
\end{figure}

Let now $m=n=7$ and consider the quasi-pattern $J$ depicted in Fig.~\ref{fig-pattern-77}.
This quasi-pattern provides the partial incidence coloring of $T_{7,7}$
given in Fig.~\ref{fig-77}, where incidences with modified colors are
in boxes and uncolored incidences are denoted
by $x$ and $y$.
Observe now that the incidences denoted by $y$ have five forbidden
colors while the incidences denoted by $x$ have four forbidden
colors.
Therefore, this partial coloring can be extended to an incidence
6-coloring of $T_{7,7}$.
\end{proof}

By Corollary~\ref{square5-corollary} and Lemmas~\ref{lemma-3k}-\ref{lemma-77},
Theorem~\ref{SW-Theorem} follows directly.

\end{document}